\documentclass[12pt]{spieman}
\usepackage{graphicx}
\usepackage{amsmath} 
\usepackage{amssymb}
\usepackage{mathrsfs}
\usepackage{setspace}
\usepackage{tocloft}
\usepackage[binary-units=true]{siunitx}
\usepackage[version=3]{mhchem}

\DeclareSIUnit\Molar{\textsc{m}}
\DeclareSIUnit\bits{\text{bits}}

\newcommand{\paren}[1]{\left(#1\right)}
\newcommand{\brac}[1]{\left[#1\right]}

\newcommand{\bal}{\begin{align*}}
\newcommand{\eal}{\end{align*}}
\newcommand{\bpm}{\begin{pmatrix}}
\newcommand{\epm}{\end{pmatrix}}

\newcommand{\balign}{\begin{align*}}
\newcommand{\ealign}{\end{align*}}

\newcommand{\eq}[2]{\begin{equation} #2 \label{#1} \end{equation}}

\title{Multiplexed Neural Recording Down a Single Optical Fiber via Optical Reflectometry with Capacitive Signal Enhancement}
\author{Samuel Rodriques$^*$, Adam Marblestone$^*$, Max Mankin, Lowell Wood, Edward Boyden}
\author{Samuel G. Rodriques,\supscr{a,b,*} Adam H. Marblestone,\supscr{a,*} Max Mankin,\supscr{c} Lowell Wood, Edward Boyden\supscr{d}}

\affiliation{\supscrsm{a}MIT Media Lab\\
\supscrsm{b}MIT Department of Physics\\
\supscrsm{c}Harvard University Department of Chemistry\\
\supscrsm{d}MIT Media Lab and McGovern Institute, Departments of Brain and Cognitive Science and Biological Engineering}

\cftpagenumbersoff{figure}
\cftpagenumbersoff{table} 

\begin{document}
\maketitle

\begin{abstract}

We introduce a fiber-optic architecture for neural recording without contrast agents, and study its properties theoretically. Our sensor design is inspired by electrooptic modulators, which modulate the refractive index of a waveguide by applying an electric field across an electrooptic core material, and allows recording of the activities of individual neurons located at points along a \SI{10}{\centi\meter} length of optical fiber with \SI{20}{\micro\meter} axial resolution, sensitivity down to \SI{100}{\micro\volt} and a dynamic range of up to \SI{1}{\volt} using commercially available optical reflectometers as readout devices. A key concept of the design is the ability to create an ``intensified'' electric field inside an optical waveguide by applying the extracellular voltage from a neural spike over a nanoscopic distance. Implementing this concept requires the use of ultrathin high-dielectric capacitor layers. If suitable materials can be found -- possessing favorable properties with respect to toxicity, ohmic junctions, and surface capacitance -- then such sensing fibers could, in principle, be scaled down to few-micron cross-sections for minimally invasive neural interfacing. Custom-designed multi-material optical fibers, probed using a reflectometric readout, may therefore provide a powerful platform for neural sensing.
\end{abstract}

\keywords{reflectometry, neural recording, fiber-optic, electrooptic, nanophotonics}

{\noindent \footnotesize{\bf Correspondence to}: \linkable{esb@media.mit.edu}\\ 
\supscrsm{*}These authors contributed equally to this work.}

\begin{spacing}{2}

\section{Introduction}\label{intro}

\paragraph{} The extracellular electrode is a classic neural recording technology. The electrode is essentially a conductive wire, insulated except at its tip, placed in the extracellular medium within a few hundred microns of a neuron of interest, where it samples the local voltage relative to a ground lead~\cite{buzsaki2015tools}. This voltage differential, typically on the order of \SI{100}{\micro\volt} in response to an action potential from a nearby neuron~\cite{marblestone2013physical}, is then amplified and digitized.

\paragraph{} The virtues of the electrode are twofold. First, the technique can reach single neuron precision by virtue of the electrode being inserted close to the measured neuron. Second, no exogenous contrast agents (i.e., genetically encoded fluorescent proteins, voltage sensitive nanoparticles, chemical dyes) are necessary: the endogenously generated electric currents in the brain are sensed directly in the form of a voltage. Ideally, for a neurotechnology to be medically valuable for a large number of human patients in a reasonably short timescale, it should not require modification of the neuron.

\paragraph{} Yet, while multi-electrode arrays allow the insertion of many electrodes into a brain, electrodes have limitations~\cite{marblestone2013physical} in scaling to the simultaneous observation of large numbers of neurons. The electrical Johnson noise scales as $R^{\frac{1}{2}}$, leading to voltage noise levels on the order of \SI{10}{\micro\volt} in standard practice. Moreover, the bandwidth of an electrical wire is limited by the cross-sectional area of the wire, due to capacitance. These factors pose limits on the scalability of electrode-based recording, as described further in \textbf{Section \ref{comparisonsection}}.

\paragraph{} In order to maintain the advantages of electrodes -- single neuron precision based on endogenous neural signals -- while enabling improved scaling performance, we turn to photonics. Telecommunications has moved from electrical to optical data transmission because of the high bandwidths and low power losses enabled by optics in comparison to electrical conductors~\cite{hilbert2011world}; the same may be helpful for neural readout technologies. Because optical radiation heats brain tissue and scatters off tissue inhomogeneities, a wired (i.e., fiber or waveguide based) optical solution may be desirable, i.e., using optical fibers to guide light so that it need not travel through the tissue itself. Second, to minimize volume displacement, signals from many neurons should be multiplexed into each optical fiber. Third, ideally the sensing mechanism would rely only on endogenous signals, e.g., electrical, magnetic or acoustic fields from the firing neurons, rather than imposing a need for exogenous protein or nanoparticle contrast agents. With on the order of 100,000 neurons per \si{\milli\meter\cubed} in the cortex, or a median spacing of roughly one neuron per cube of size \SI{21.5}{\micro\meter}, we require an axial resolution of sensing in the range of \SI{20}{\micro\meter}. Note that every neuron in a mammalian brain is within a few tens of microns of the nearest capillary~\cite{blinder2013cortical}, well within the distance necessary for direct electrical sensing of the action potential~\cite{marblestone2013physical}, and thus in principle the fine microvessels of the cerebral vasculature could serve as a delivery route for neural activity sensors, if the fibers could be made sufficiently thin~\cite{llinas2005neuro}, i.e, well below \SI{10}{\micro\meter} for the smallest capillaries. Thus, the system should be compatible with a variety of form factors, e.g., thin flexible fibers suitable for minimally invasive endovascular delivery~\cite{bower2013intravenous, llinas2005neuro}, or rigid pillars suitable for direct penetration of the brain parenchyma~\cite{du2011multiplexed}. 

\paragraph{} Our proposed architecture is based on two powerful technologies developed by the photonics industry: fiber optic reflectometry, which enables optical fibers to act as distributed sensors~\cite{jarzynski1987fiber,Sorin1993,palmieri2013distributed,bao2012recent}, and electrooptic modulators based on the plasma dispersion effect, which generate large changes in the index of refraction of a waveguide in response to relatively small applied voltages~\cite{Liu2007high,Liu2004high,xu2005micrometre,soref2006past,jalali2006silicon}. By combining reflectometry with electro-optic modulation, we demonstrate that it would be possible to do spatially multiplexed neural recording in a single optical fiber.

\section{Design Principles}

\begin{figure}
\begin{center}
\begin{tabular}{c}
\includegraphics[width=\textwidth]{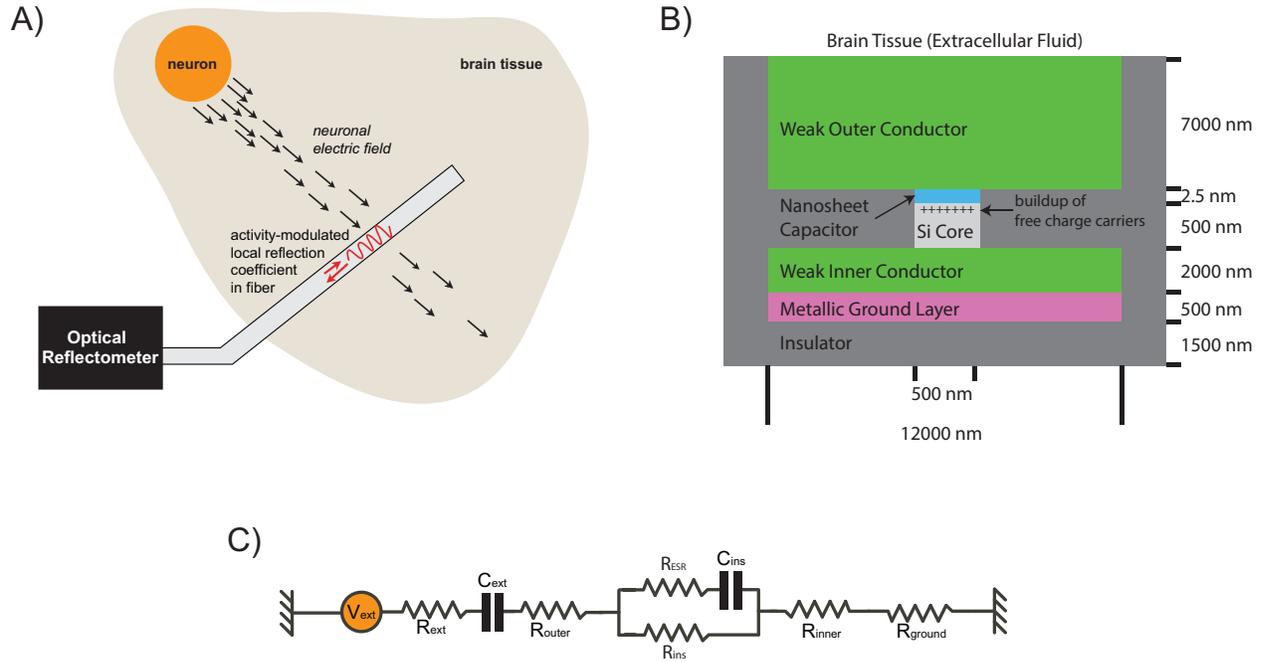}
\end{tabular}
\end{center}
\caption{\label{Fig1} \textbf{A) High-level architecture.} An optical fiber inserted into the brain acts as a distributed sensor for neuronal activity, which is read out reflectometrically. \textbf{B) Structure of the reflectometric probe.} When a voltage is applied across the nanosheet capacitor insulator layer, free holes in the inner conductor and in the silicon build up on the surface of the insulator layer and alter the refractive index. In order to magnify number of charge carriers accumulated on the surface of the insulator layer, an insulating film with an extremely large dielectric constant is used. \textbf{C) Circuit diagram of the device.} The circuit diagram of the device consists of a resistor representing each of the material layers between the neuron and ground, and two capacitors, one of which ($C_\text{ext}$) represents the interfacial capacitance and the other of which ($C_\text{ins}$) represents the capacitance of the insulating layer. The effective series resistance $R_\text{ESR}$ of the insulating region capacitor can be ignored provided the insulating region has dielectric loss $\tan(\delta) \ll 1$, and the parallel resistance of the insulating layer $R_\text{ins}$ can be ignored provided it is much larger than $R_\text{ground}$. If in addition $R_\text{ground}$ is chosen to be larger than the other resistances in the circuit, the capacitances $C_\text{ins}$ and $C_\text{ext}$ may be treated as series capacitances (see text).}
\end{figure}

Reflectometers are capable of measuring changes in the index of refraction along the length of an optical fiber by sending optical pulses down the length of the fiber and recording the times and magnitudes of returning reflections~\cite{Sorin1993}. We propose to use reflectometry to sense neural activity at many points along the length of an optical fiber, as shown in figure~\ref{Fig1}A. The local voltage at a given position along the fiber would modulate its local index of refraction via the free carrier dispersion effect, giving rise to reflections. A reflectometer located outside the brain would then determine, at each time, the spatial profile of extracellular voltage along the length of the fiber.

\subsection{Fiber-Optic Reflectometry}

The sensitivity of a reflectometer is given in terms of fractional power reflected relative to the input power, measured in decibels. To determine the magnitude of the reflections generated by a change in local refractive index inside a fiber, note that when an electromagnetic plane wave propagates in a material with refractive index $n_1$, and is normally incident on a material with refractive index $n_2$, the intensity reflected is given by the Fresnel equation
\eq{}{R = \paren{\frac{n_1 - n_2}{n_2 + n_1}}^2.}
Assuming small changes in $n$, we can approximate $R$ by
\eq{DeltaR}{R = \paren{\frac{\Delta n}{2 n}}^2.}
Expressed in decibels, the magnitude of the reflections generated by the change in refractive index is
\eq{}{10 \log_{10}(R).}

Below, we will assume the use of reflectometers that can achieve \SI{20}{\micro\meter} spatial resolution and sensitivities down to \SI{-130}{\decibel} (corresponding to $R\sim 10^{-13}$) with \SI{12}{\hertz} repetition rates over \SI{8.5}{\meter}. This corresponds to a measurement time of \SI{1}{\milli\second} for any given \SI{10}{\centi\meter} segment of fiber, so using a similar device we anticipate that it would be possible to sense reflections along the length of a \SI{10}{\centi\meter} fiber with a repetition rate of \SI{1}{\kilo\hertz}, \SI{20}{\micro\meter} spatial resolution and sensitivity of \SI{-130}{\decibel}. 

For example, the commercially available Luna OBR 5T-50 and ODiSI B systems exhibit an optical noise floor of about $-120$ to $-125$ dB and a dynamic range of $-60$ dB, while the OBR 4600 gives a $<-130$ dB optical noise floor. Optical phase noise associated with the laser is limiting in such optical frequency domain reflectometry (OFDR) systems~\cite{soller2005high}; Littman-Metcalf external cavity tunable lasers, with narrow linewidths and low phase noise, can be swept at \SI{1}{\kilo\hertz} repetition rates over an optical frequency range of several \si{\tera\hertz}, leading to an OFDR spatial resolution of roughly \SI{20}{\micro\meter}.

If additional speed is required, even faster reflectometers are available that achieve similar sensitivities; one reflectometer has been demonstrated that achieves \SI{-130}{\decibel} sensitivity over \SI{2}{\kilo\meter} with millimeter spatial resolution and a \SI{10}{\hertz} repetition rate~\cite{gifford2007millimeter}, corresponding to a sensing rate of $2\times 10^7$ measurements per second, a factor of 4 improvement over the commercial reflectometer mentioned above. If additional sensitivity is needed, reflectometers can achieve sensitivities as low as \SI{-148}{\decibel}~\cite{sorin1992measurement}, while systems using fiber-based optical amplifiers and super-fluorescent fiber sources can achieve sensitivities as low as \SI{-160}{\decibel}, which is nearly shot-noise limited~\cite{Takada1993}.

\subsection{Electro-optic modulation}

Silicon electro-optic modulators are widely used in photonics to alter the propagation of light through a material in response to an applied voltage~\cite{lipson2005guiding,soref2006past}. Typical applications of electrooptic modulators take the form of electrically controlled optical switches: signals on the order of \SI{5}{\volt} are used to drive optical phase shifts on the order of $\pi$. These devices are optimized for GHz bandwidths, with the goal of providing high speed, low power microchip interconnects~\cite{xu2005micrometre}, with bandwidths up to \SI{30}{\giga\hertz} possible~\cite{phare201430}. Here, however, we are interested in the application of similar device physics to a very different problem: sensing extracellular neuronal voltages on the order of \SI{100}{\micro\volt} at \SI{1}{\kilo\hertz} rates. Thus, our required switching rate is 1 millionfold slower, yet our required sensitivity is on the order of 1 millionfold better. We are thus concerned with the design of electrooptic modulators optimized for sensitivity rather than bandwidth.

\subsubsection{Insufficiency of endogenous electric fields}

The voltages present in the extracellular medium due to neural activity range from \SIrange{10}{1000}{\micro\volt}, with electric fields on the order of \SI{1}{\volt\per\meter}~\cite{marblestone2013physical}. These electric field arise from the fact that the extracellular voltage from a spiking neuron decays over a distance on the order of \SI{100}{\micro\meter}. Note that the \emph{transmembrane} voltage during an action potential is much larger, on the order of \SI{100}{\milli\volt}.

Although it is possible to measure electric fields down to microvolts per meter using refractive index sensors, such sensors rely on very long sensing distances and can only measure the average field over the length of the fiber~\cite{vohra1992fiber,jarzynski1987fiber,ku1983high,donalds1982electric}. For measurements that preserve spatial information, sensitivities down to hundreds of volts per meter have been obtained using the Stark shift~\cite{Czarnetzki2007ultra,takizawa2004sensitive,takizawa2002observation,czarnetzki1998sensitive}. Thus, the endogenous electric fields in the brain are too small to be \emph{directly} read out using such physical coupling strategies in conjunction with optical reflectometers.

For sensing, it will therefore be necessary to create an \emph{intensified} electric field $E = V/d$ by causing the extracellular voltage $V$ to drop over a nanoscopically narrow region of insulating material of thickness $d$. This intensified electric field could then in principle be detected electrooptically. 

\subsubsection{The free carrier dispersion effect}

The design shown in figure~\ref{Fig1}B consists of an extended multi-layer semiconductor waveguide on a grounded metal substrate, surrounded on three sides by insulation and on the fourth side by brain tissue or extracellular fluid. The ``outer conductor,'' ``inner conductor,'' and ``silicon'' core layers are weak conductors which function as resistive layers between the brain and ground. Between the core and outer conductor there is an ultra-thin insulator layer that functions as a capacitor with large capacitance, over which most of the voltage drops. Light propagates through the silicon waveguide core, which is doped to achieve sufficiently high conductivity to allow most of the voltage to drop over the insulator layer. Note that the inner conductor is chosen to be thick enough to prevent optical attenuation due to the metal substrate (although there are other possible methods to reduce attenuation due to the metal, e.g., by removing the metal from the region directly under the waveguide, as in~\cite{Liu2004high}), and the metal substrate is chosen thick enough to provide a high-fidelity ground throughout the fiber. Clearly, the index of refraction of the inner and outer conductors must be smaller than that of the silicon core, which is assumed to be around 3 at telecommunications frequencies.

The design relies on the free carrier dispersion effect (also known as the plasma dispersion effect) in silicon: the index of refraction within the silicon core changes due to the injection of charge carriers into the silicon when an electric field is applied across the thin insulator layer~\cite{soref1987,soref1987b,Liu2004high}. In addition to the free carrier effect, there exist other modalities of electrooptic modulation, such as the linear electrooptic (Pockels) effect~\cite{dalton2009electric}, the quadratic electrooptic effect (Kerr) effect~\cite{Shen2014} and the Stark effect~\cite{Czarnetzki2007ultra}. All of these effects benefit from reducing the thickness $d$ of the insulator layer to create a large electric field $V/d$~\cite{adams1986optimum}. However, the magnitude of the free carrier dispersion effect may be increased further by increasing the dielectric constant of the insulator layer. For a material with a suitably large value of $\epsilon/d$, the change in refractive index due to the free carrier effect will be much larger than the changes that can be obtained via the other electrooptic effects. Many current integrated semiconductor electrooptic modulators are based on the free carrier dispersion effect~\cite{Liu2004high,Liu2007high}.

Changes in the index of refraction in the free carrier modulated sub-layer of the silicon may be modeled as changes in the overall effective index of refraction of the fiber~\cite{paul1987fiber}. The magnitude of this effective change is given by weighting the magnitude of the change in the free carrier modulated layer by the percentage of power contained in that layer, i.e.,
\eq{}{\Delta n_\text{eff} = (1-\eta) \Delta n_\text{active},}
where $n_\text{active}$ is the index of refraction in the free carrier modulated layer and $1-\eta$ is the fraction of the power in the beam contained in the active region. 

We will denote by $d$ the thickness of the insulator layer, by $b$ the thickness of the layer of injected charge carriers in the silicon, and by $a$ the remaining thickness of the silicon layer. 
An order-of-magnitude approximation for $\eta$ is then given by
\eq{etadef}{\eta \cong \frac{a}{(a + b)}}
and we have
\eq{Neff}{\Delta n_\text{eff} = \paren{1 - \frac{a}{(a + b)}} \Delta n_\text{active}.}
For this reason, the silicon waveguide is chosen to be thin to maximize the percentage of the optical wave contained in the layer containing the injected charges. Because of the deep subwavelength thickness of the active layer, a precise calculation of $\Delta n_\text{eff}$ could be done using a full vectorial Maxwell simulation of the waveguide modes~\cite{Liu2004high}, but for our purposes the approximation of equation \ref{Neff} suffices to illustrate the basic scaling.

Upon applying a voltage across the insulating layer capacitor, the density of charge carriers injected into the active layer inside the silicon core, denoted by $\Delta c$, is simply given by the equation for capacitance:

\eq{deltac}{\Delta c = \frac{C_\text{ins}}{A b} \Delta V_\text{ins},}
where $C_\text{ins}$ is the capacitance of the insulating region, $A$ is the area over which the charges are distributed, $b$ is the thickness of the layer of injected charge carriers in the silicon, and $\Delta V_\text{ins}$ is the voltage dropped over the insulating region. Equation~\eqref{deltac} may be recast in terms of the total voltage $\Delta V$ applied over the device by introducing an effective capacitance $C_\text{eff}$, such that 
\eq{}{\Delta c = \frac{C_\text{eff}}{A b} \Delta V.}
In practice, $C_\text{eff}$ will only deviate significantly from $C_\text{ins}$ when the capacitance of the brain-fiber interface is significant (discussed below). The change in refractive index in the region with the injected charge is related to the change in the carrier concentration by a power law~\cite{soref1987b}. When the injected carriers are holes, the magnitude of the electrooptic effect is greatest. The relation is then, for the change in refractive index in the injected charge region,
\eq{HoleRefIndChange}{\Delta n_\text{active,h} = C_h \brac{\frac{C_\text{eff}}{A b} \Delta V}^{0.8}} where $C_h$ is an empirically defined constant.

For silicon, the value of $C_h$ is given by~\cite{Liu2004high}, for $\SI{1.55}{\micro\meter}$ light,
\eq{}{C_h = \SI{-8.5e-18}{\centi\meter^{2.4}}.}

Similar values are obtained for other semiconductors and other wavelengths~\cite{Bennett1990,soref1987b}. To find the effective refractive index within the silicon waveguide, we multiply equation~\eqref{HoleRefIndChange} by the volume factor $1-\eta$ from equation~\eqref{etadef}. Assuming $b \ll a$ (i.e., that the injected charge layer is deeply sub-wavelength while the waveguide core thickness is on the same order as the wavelength), we find 
\eq{RefIndChangeHoles}{\Delta n_\text{eff} \cong C_h \frac{b^{0.2}}{a}\brac{\frac{C_\text{eff}}{A} \Delta V}^{0.8}.}
Since $b$ will be on the order of \SI{10}{\nano\meter}, we will henceforth assume $b^{0.2} \approx \SI{0.01}{\meter^{0.2}}$.

\subsubsection{Effect of the brain-electrode interface}

The brain-electrode interface also has a finite capacitance $C_\text{ext}$ of approximately \SI{1}{\farad\per\square\meter}~\cite{franks2005impedance,arfin2006miniature}, which arises due to the presence of a double-layer and which can be calculated via the Gouy-Chapman equation. Figure~\ref{Fig1}C shows a circuit diagram of the device which includes this interfacial capacitance. At \SI{<1000}{\hertz}, and subject to appropriate materials choices (see section \ref{MatProps}), the impedance of the circuit is dominated by the two capacitors $C_\text{ins}$ and $C_\text{ext}$ rather than by purely resistive elements of the circuit. For this reason, we may ignore the purely resistive elements and treat the two capacitances as though they were in series. To a good approximation, therefore, the charge that accumulates on the surface of the insulating region in response to a voltage $\Delta V$ across the entire device is given by
\eq{}{Q = C_\text{eff} \Delta V,}
where the effective capacitance $C_\text{eff}$ of the surface and insulating region capacitors in series is
\eq{Ceff}{C_\text{eff} = \frac{1}{\frac{1}{C_\text{ins}}+\frac{1}{C_\text{ext}}}.}
The capacitance of the insulating region is given by
\eq{}{C_\text{ins} = \epsilon_0 A \frac{\epsilon}{d},}
where $d$ is the thickness of the insulating region, $\epsilon$ is the relative permittivity, $A$ is the sensing area of the device, and $\epsilon_0$ is the permittivity of free space. The capacitance per unit area is specified by $\epsilon/d$, which will be the primary figure of merit for determining the sensitivity, noise, and dynamic range of the device.

The effective capacitance $C_\text{eff}$ is shown in figure~\ref{ParametrizationFig}A as a function of $\epsilon/d$, assuming a surface capacitance per unit area of approximately \SI{1}{\farad\per\square\meter}~\cite{franks2005impedance,arfin2006miniature} and a sensing length of \SI{20}{\micro\meter}. Note that because a pair of series capacitors is dominated by the smaller capacitor, the silicon waveguide core is kept only \SI{500}{\nano\meter} wide in order to ensure that the capacitance per unit length of the insulating layer remains smaller than the capacitance per unit length of the \SI{12}{\micro\meter}-wide brain-fiber interface.

\subsubsection{Noise}

The RMS amplitude of the Johnson noise in the voltage across the capacitor is given by
\eq{}{V_\text{RMS} = \sqrt{4 k_B T R_c \Delta F}}
where $\Delta F = 1/(2 \pi R_c C_\text{eff})$ and $R_c$ is the total resistance governing charging/discharging of the capacitor. Hence
\eq{JohnsonNoise}{V_\text{RMS} \sim \sqrt{k_B T/C_\text{eff}}}
which does not depend on the resistance $R_c$.

\subsubsection{Dynamic range}

In addition to the constraints on the minimum resolvable voltage described above, the maximum resolvable voltage is limited by the total number of charge carriers available in the outer conductor layer. In particular, if a voltage $V$ is applied over the device capacitance $C_\text{eff}$, a number of charge carriers equal to $V C_\text{eff} / e$ must be recruited to the surface of the insulator layer, where $e$ is the electron charge. Hence, if the outer conductor has a volume of $\mathcal{V}$ associated with each recording site and a free charge carrier density equal to $c$, the saturation voltage is given by
\eq{DynamicRange}{V_\text{max} = \frac{c \mathcal{V} e}{C_\text{eff}}.}

For this reason, the outer conductor layer is chosen to occupy most of the thickness of the fiber in order to maximize the dynamic range.

\section{Predicted properties}

\begin{figure}
\begin{center}
\begin{tabular}{c}
\includegraphics[height=0.75\textheight]{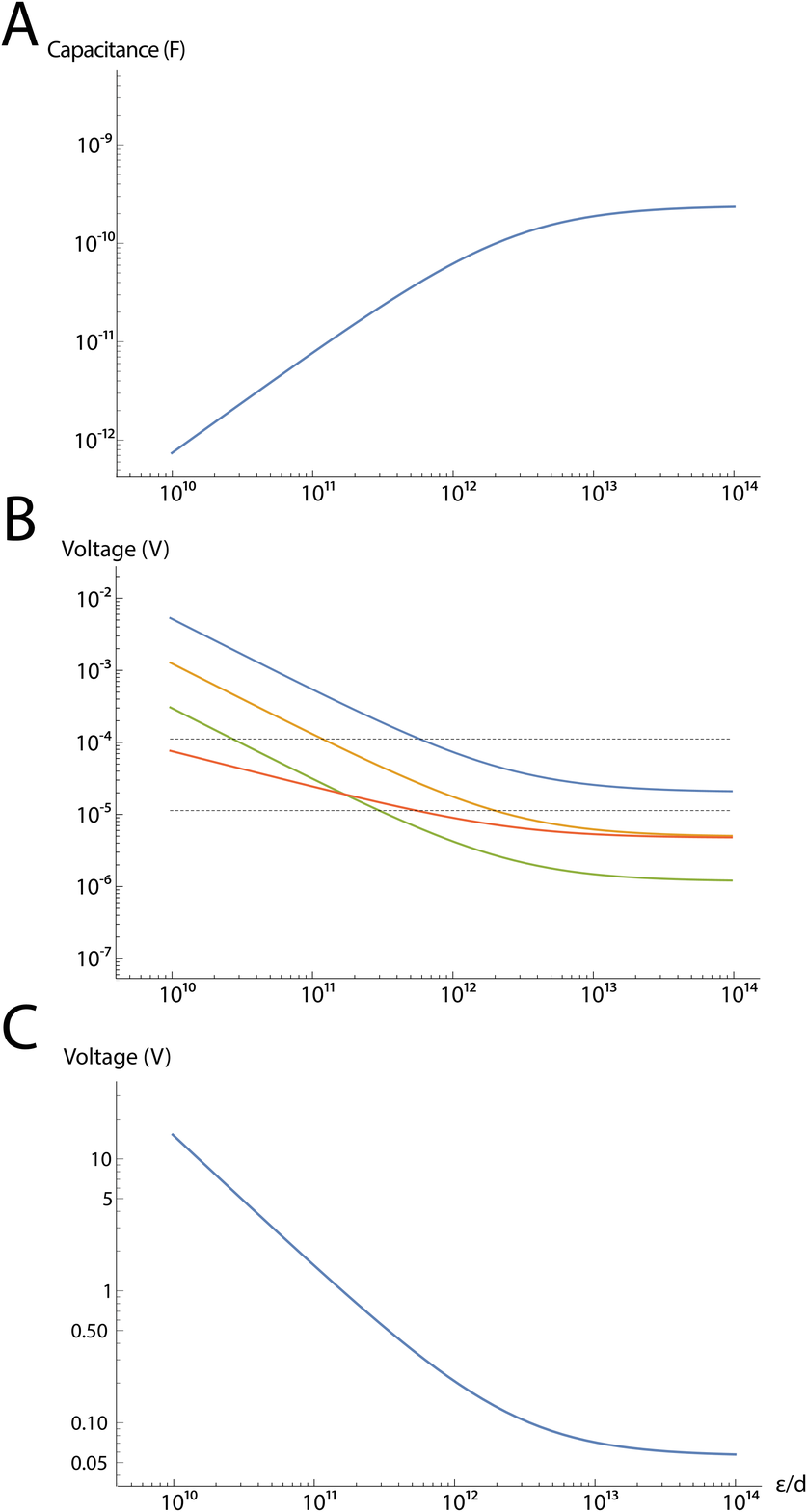}
\end{tabular}
\end{center}
\caption{\label{ParametrizationFig} \textbf{Properties of the design parametrized by $\epsilon/d$.} A) The effective capacitance $C_\text{eff}$ given in equation~\eqref{Ceff} is shown as a function of $\epsilon/d$ assuming the dimensions shown in figure~\ref{Fig1}, and a capacitance of \SI{1}{\farad\per\square\meter} at the surface of the device. B) The change in voltage required to give rise to a reflection at the \SI{-120}{\decibel} level (blue), the \SI{-130}{\decibel} level (orange), and the \SI{-140}{\decibel} level (green) is shown as function of $\epsilon/d$, obtained by solving equation~\eqref{RefIndChangeHoles} for $\Delta V$. Also shown is the magnitude of the Johnson noise given in equation~\eqref{JohnsonNoise} as a function of $\epsilon/d$ (red). The black dashed lines correspond to \SI{10}{\micro\volt} and \SI{100}{\micro\volt}. For example, detecting \SI{100}{\micro\volt} signals at the \SI{-130}{\decibel} level requires a figure of merit $\epsilon/d > 10^{11}$.  C) The saturation voltage $V_\text{max}$ given in equation~\eqref{DynamicRange} is shown as a function of $\epsilon/d$ assuming a carrier concentration of \SI{5e16}{\per\cubic\centi\meter} in the outer conductor.}
\end{figure}

The key figure of merit determining the properties of the device is $\epsilon/d$, where $\epsilon$ is the dielectric constant of the insulating layer and $d$ is the thickness of the insulating layer. The figure of merit is proportional to the capacitance per unit area of the insulating region, and it determines the sensitivity, noise, and dynamic range via equations~\eqref{RefIndChangeHoles},~\eqref{JohnsonNoise}, and~\eqref{DynamicRange}. In figure~\ref{ParametrizationFig}B, the sensitivity of the device is shown as a function of $\epsilon/d$ for devices of width \SI{12}{\micro\meter}. The blue, orange, and green lines correspond to the voltages which would be sensed at \SI{-120}{\decibel}, \SI{-130}{\decibel}, and \SI{-140}{\decibel} respectively at the given value of $\epsilon/d$. The red line corresponds to the level of the Johnson noise at the given value of $\epsilon/d$. The black dashed lines indicate the \SI{10}{\micro\volt} and \SI{100}{\micro\volt} ranges. The power law region (a straight line on the log-log plot) corresponds to the region in which $C_\text{eff} \approx C_\text{ins}$, so that the reflection coefficient $R \propto (\epsilon \Delta V/d)^{0.8}$. For values of $\epsilon/d$ much greater than \SI{e12}{\per\meter}, we have $C_\text{eff} \approx C_\text{ext}$, so the sensitivity does not improve with increasing $\epsilon/d$.

Clearly, in order to be able to sense a \SI{100}{\micro\volt} signal using a reflectometer with \SI{-130}{\decibel} sensitivity, a value of $\epsilon/d > \SI{2e11}{\per\meter}$ is required. With a reflectometer of sensitivity \SI{-140}{\decibel}, a \SI{100}{\micro\volt} signal can be sensed with a value of $\epsilon/d$ as low as \SI{4e10}{\per\meter}. However, even with a reflectometer of sensitivity \SI{-150}{\decibel}, a value of $\epsilon/d > \SI{e10}{\per\meter}$ would be required to overcome Johnson noise. Finally, if a value of $\epsilon/d$ on the order of \SI{2e12}{\per\meter} could be obtained, it would be possible to detect signals as small as \SI{10}{\micro\volt} with commercially available reflectometers with \SI{-130}{\decibel} sensitivity.

\subsection{Required material properties}\label{MatProps} 

The primary electrical constraint on the device is that the impedance at \SI{1}{\kilo\hertz} must be dominated by the insulator capacitance. If the capacitance per unit area of the capacitor is \SI{0.1}{\farad\per\meter}, corresponding to a value of $\epsilon/d$ equal to \SI{e10}{\per\meter}, then the capacitance of a region with width \SI{12}{\micro\meter} and length \SI{20}{\micro\meter} is \SI{24}{\pico\farad}, corresponding to an impedance of \SI{4e7}{\ohm} at \SI{1}{\kilo\hertz}. Assuming that the metal layer has a resistivity no greater than \SI{100}{\nano\ohm\meter} (10x that of silver), if the metal layer is made at least \SI{500}{\nano\meter} thick, it will have a resistance of \SI{10000}{\ohm} along the entire length of the fiber, which is much less than that of a single sensing region of the capacitor, and which is also large enough for the input impedance of an implanted recording device~\cite{cogan2008neural}. Moreover, in this case, the $RC$-time constant of the system consisting of the metal ground and the effective capacitor (for a single sensing region) will be \SI{240}{\nano\second}, which is much less than \SI{1}{\milli\second}. The resistance of the inner conductor, outer conductor, and core will be negligible compared to the huge capacitive impedance, provided they are chosen to be semiconductors. The inner conductor must be chosen to make ohmic junctions with the metal and silicon layers.

\paragraph{} In addition, the outer conductor is constrained by the requirement that its conductivity be less than or on the order of that of the extracellular medium (approximately \SIrange{0.1}{0.3}{\siemens\per\meter}~\cite{ranck1963specific,nicholson1965specific}) to avoid lateral propagation of the electrical signals from the brain along the length of the fiber. Finally, to maximize the saturation voltage, it is necessary to choose the outer conductor to have a large free carrier density. These two properties can be satisfied simultaneously using a semiconductor with very low mobility. As an example, nickel oxide is a p-type semiconductor known to have a mobility between \SI{0.1}{\square\centi\meter\per\volt\per\second} and \SI{1}{\square\centi\meter\per\volt\per\second}, but with a conductivity of \SIrange{0.1}{1}{\siemens\per\meter} in the undoped state~\cite{chen2005characterization,liu2014nickel}, corresponding to a free carrier concentration on the order of \SI{5e16}{\per\cubic\centi\meter}, which is ideal for our application. With these material properties, the saturation voltage can be calculated from equation~\eqref{DynamicRange} as a function of the figure of merit $\epsilon/d$, and is shown in figure~\ref{ParametrizationFig}C. Alternatively, films of semiconducting nanoparticles may provide similarly large carrier concentrations at low mobilities, and may be significantly simpler to fabricate~\cite{klem2008impact}.

\subsection{Comparison of electrical and optical performances}\label{comparisonsection}

The bandwidth of an electrical wire is limited by its cross-sectional area by 
\eq{BWRequirement}{\text{BW} \approx \frac{A \log(1+\alpha)}{(1+\alpha)^2 4 \pi^2 \epsilon_r \epsilon_0 \rho L^2}}
where $\rho$ is the resistivity of the wire, $L$ its length, $A$ its cross-sectional area, $\alpha$ is the thickness of the insulation as a fraction of the thickness of the wire, $\epsilon_r$ is the dielectric constant of the medium surrounding the conductor and $\epsilon_0$ is the permittivity of free space. For any given value of the bandwidth, equation~\eqref{BWRequirement} specifies a minimum total cross-sectional conductor area that must be available to convey the information. Note that, according to the Shannon-Hartley theorem, the bandwidth required to achieve an information rate $I$ is given in terms of the signal power $S$ and noise power $N$ by 
\eq{SHtheorem}{\text{BW} = \frac{I}{\log_2(1+S/N)}.}

Optical waves can be spatially confined to channels with thickness on the order of the wavelength of light (or below, e.g., using high-index-contrast polymer fibers or nanoslot waveguides~\cite{hochberg2007towards}), yet can transmit data at rates of terabits per second over long distances. To be specific, an information rate of roughly $I = 10^4$ bits per second per neuron is required for analog neural recording. At a signal to noise ratio of 10, sensing 1000 neurons requires a bandwidth of roughly \SI{3e7}{\hertz} via equation~\eqref{SHtheorem}. In theory, this is trivial even for an optical channel of \SI{500}{\nano\meter} width and \SI{10}{\centi\meter} length; yet for an electrical channel consisting of a \SI{500}{\nano\meter} wide electrical conductor, \SI{500}{\nano\meter} insulation and \SI{10}{\centi\meter} length, this is already at the limit of achievable bandwidth as calculated from equations~\eqref{BWRequirement} and~\eqref{SHtheorem}. In practice, the field of electrical recording has advanced greatly in recent years, and efforts are underway to package $\sim 1000$ electrical recording sites into shanks with \SI{50}{\micro\meter} width~\cite{buzsaki2015tools,scholvin2015close}. However, our calculations suggest that with further miniaturization, optical methods will surpass the fundamental limits of wired electrodes in terms of space efficiency. Thus, improvements in nanophotonic miniaturization could become a key enabler for the comprehensive optical readout of neural circuit dynamics.


\section{Discussion}

Ultra-large-scale neural recording is highly constrained both by physics and by the biology of the brain~\cite{marblestone2013physical}. Here we have argued that an architecture for scalable neural recording could combine a) the use of optical rather than electronic signal transmission to maximize bandwidth, b) confined rather than free space optics to obviate the effects of light scattering and absorption in the tissue, c) spatial or wavelength multiplexing within each optical fiber in order to minimize total tissue volume displacement, d) a thin form factor to enable potential deployment of fibers via the cerebral vasculature, e) direct electrical sensing to remove the need for exogenous dyes or for genetically encoded contrast agents. 

In our proposed design, the \SI{100}{\micro\volt} scale extracellular voltage resulting from a neuronal spike is applied across a nanoscopically thin, high-dielectric capacitor. Charging of the capacitor results in the buildup of free charge carriers in the neighboring silicon waveguide core, altering the local refractive index of the silicon and causing a detectable optical reflection. Reflectometry then enables multiplexed readout of these spike-induced reflections.

A key challenge in implementing such a design is to achieve a figure of merit $\epsilon/d$ for the capacitor in excess of $10^{11}$. While current nanosheet capacitors~\cite{kim20142d, wang2014all, ngo2014epitaxial} approach $\epsilon/d \approx 10^{11}$, achieving sufficiently high figure of merit in a practical fabrication process may be difficult. Alternatively, there exist ultra-high-dielectric ($\epsilon > 10^5$) materials such as calcium copper titanate~\cite{adams2002giant}, but much of their dielectric behavior derives from boundaries between microscopic crystal grains, and whether these materials can maintain their dielectric properties in sufficiently thin nanoscopic layers is unclear. Another challenge is to maintain high carrier concentrations in the outer conductor without creating high electron mobilities which could lead to attenuation of the optical wave propagating in the nearby core. The $\SI{12}{\micro\meter}$ thickness of the device presented here is likewise limited by material concerns. If the carrier concentration of the outer conductor could be increased while maintaining low mobility, further miniaturization of the device would be possible.

If appropriate materials combinations can be fabricated, we have shown that the device could achieve the requisite sensitivity, noise level, dynamic range and response time for recording both neural spikes and local field potentials. More broadly, our results suggest that integrated photonics could enable highly multiplexed readout of neuronal electrical signals via purely optical channels.

\acknowledgments
We thank Jorg Scholvin, Michal Lipson, Mian Zhang, Chris Phare, Konrad Kording, Dario Amodei, Deblina Sarkar and George Church for helpful discussions. Funding was contributed by the Fannie and John Hertz Foundation (SR, AM, MM), the NIH Director's Pioneer Award (ESB) and NIH grants 1U01MH106011 (ESB) and 1R24MH106075-01 (ESB, AM). All code will be posted at \\\linkable{http://scalablephysiology.org/fiber} for use by the community. 

\bibliographystyle{spiejour} 
\bibliography{./fiber-reflectometry}

\end{spacing}
\end{document}